\documentstyle[prl,aps]{revtex}
\begin{document}

\input epsf
\draft
\twocolumn[\hsize\textwidth\columnwidth\hsize\csname
@twocolumnfalse\endcsname                             
\title{Vortex matter in superconducting mesoscopic disks: Structure,
magnetization, and phase transitions}
\author{J. J. Palacios}
\address{Department of Physics and Astronomy, University of Kentucky,
Lexington, KY 40506, USA \\ and 
Departamento de F\'{\i}sica Te\'orica de la Materia Condensada, 
Universidad Aut\'onoma de Madrid, Cantoblanco, Madrid 28049, Spain.}

\date{\today}
\maketitle

\widetext
\begin{abstract}
\leftskip 2cm
\rightskip 2cm

The dense vortex matter structure and associated magnetization 
are calculated for type-II superconducting mesoscopic disks.
The magnetization exhibits generically first-order phase transitions 
as the number of vortices changes by one and 
presents two well-defined regimes:  A {\em non-monotonous} 
evolution of the magnitude of the magnetization
jumps signals the presence of a vortex glass structure which is 
separated by a second-order phase
transition at $H_{c2}$ from a condensed state of vortices (giant
vortex) where the magnitude of the jumps changes monotonously.
We compare our results with Hall magnetometry measurements 
by Geim  et al. (Nature {\bf 390}, 259 (1997)) and 
claim that the magnetization exhibits clear
traces of the presence of these vortex glass states.

\end{abstract}

\pacs{\leftskip 2cm PACS numbers: 74, 74.60.Ec, 74.76.-w}
\vskip2pc]

\narrowtext
Electronic device miniaturization in semiconductors 
has recently reached the ultimate  
limit where the number of electrons present in the device can
be tuned at will even down to a single electron\cite{Ashoori,Kouwenhoven}. 
These systems have been given the name of quantum dots or 
artificial atoms since their generic properties are determined 
by their few-electron configurations much the same
as in real atoms. Analogies can be drawn between these artificial atoms
and type-II superconducting mesoscopic disks in perpendicular magnetic fields 
where the role of the electron is played in this case, not by the
Cooper pair, but by another fundamental entity: The vortex. When the
dimensions of the disk are comparable to
the coherence length $\xi$ only few vortices can coexist in the system.
In contrast to the usual triangular arrangement in bulk, complex and unique 
vortex structures are expected to occur due to the competition between
surface superconductivity and vortex-vortex interaction.
When the dimensions of the system are much
smaller than $\xi$ the very notion of superconductivity 
needs to be revised\cite{grains}.

Transport experiments have contributed in a decisive way to unveil the
electronic structure of artificial atoms\cite{Kouwenhoven}. Similarly,
transport measurements\cite{Moshchalkov} in
individual mesoscopic disks gave us the first experimental evidence  of
the structure of the order parameter in these systems. Oscillations of the 
critical temperature $T_c$ as a function of the external 
magnetic field $H$ were
correctly accounted for by the quantization of the angular momentum $L$ 
of the Cooper pair wavefunction or, in other words, by transitions between
giant vortex states with a different number of flux quanta. Close
to the supercondutor-normal phase boundary the theoretical analysis of
these transitions does not present any difficulty 
since it simply implies solving the linearized Ginzburg-Landau
equations\cite{Moshchalkov,Benoist,Buzdin,Bezryadin}.

Hall magnetometry\cite{Geim}, on the other hand, is revealing itself as a
powerful tool for obtaining information of the order parameter
through its associated magnetization $M$
away from the supercondutor-normal phase boundary. To
date, the theoretical efforts to calculate the structure of the 
order parameter and $M$ well into the superconducting phase have
been mostly restricted to the numerical solving  
of the Ginzburg-Landau equations under the assumption
of an order parameter with a well-defined $L$\cite{Moshchalkov2,Peeters}. 
This is justifiable for type-I superconducting disks.
For type-II disks, however, this assumption is no longer valid. More
precisely, it is only expected to hold in the Meissner state, which is
associated to an $L=0$ order parameter, 
and above $H_{c2}$ where, close to the surface of the disk,
the superconductivity can survive up to a
higher critical field $H_{c3}$\cite{Moshchalkov,Benoist,Buzdin,Bezryadin}. 
This surface order parameter is referred to as a giant vortex or
macrovortex\cite{Moshchalkov2,Peeters}. 
For $H_{c1}<H< H_{c2}$ it has been argued\cite{Moshchalkov2} and shown in 
numerical simulations of the time-dependent Ginzburg-Landau 
equations\cite{Geim,Lopez} that the order parameter can form 
complex structures of single-fluxoid vortices, i.e., a 
``budding" Abrikosov lattice. From the more analytic standpoint
presented in this paper the appearance of these structures can only 
come about if the order parameter {\em does not} have a well-defined $L$. 
To find and understand the structure of these
vortex states, to calculate the magnetization 
associated to them, and to link these states to the Abrikosov lattice
which emerges
in the thermodynamic limit of infinite disks are the main goals of this
paper. We do this by expanding the order parameter in an appropriate basis
and minimizing, analytically to a large extent, the Ginzburg-Landau functional. 
For an increasing magnetic field we find first-order phase 
transitions whenever a vortex is added to the disk. A {\em non-monotonous}
behavior of the magnitude of the magnetization jumps signals the presence of 
the glassy vortex structures while, at $H_{c2}$, a 
second-order phase transition constitutes the signature of the condensation of
the vortices into the giant vortex state.
In addition we compare with the magnetization measured in Al
disks by Geim {\em et al.} in Ref.\ \onlinecite{Geim} and claim that the
magnetization jumps present a non-monotonous evolution which, as we
stated above and shown below, can be
associated to the existence of single-fluxoid vortex glassy structures.
This would indicate that the Al disks in Ref.\ \onlinecite{Geim} 
behave like type-II superconductors rather than 
type-I, possibly due to the expected enhancement of the 
effective magnetic penetration length in such a geometry.

We start from the traditional Ginzburg-Landau functional for the Gibbs
free energy of the superconducting state
\begin{eqnarray}
G_s&=&G_n+\int d{\bf r} \left[ \alpha |\Psi({\bf r})|^2 + 
\frac{\beta}{2}|\Psi({\bf r})|^4 + \right.\nonumber \\
&&\left.\frac{1}{2m^*}\Psi^*({\bf r})\left(-i\hbar{\bf\nabla} - 
\frac{e^*}{c}{\bf A({\bf r})}\right)^2\Psi({\bf r}) +
\frac{[h({\bf r})-H]^2}{8\pi} \right],
\label{G-L}
\end{eqnarray}
where $G_n$ is the Gibbs free energy of the normal state and $[-i
\hbar \nabla - e^* {\bf A}({\bf r})/c]^2/2m^*$ is the kinetic energy
operator for Cooper pairs of charge $e^*=2e$ and mass $m^*=2m$ in a
vector potential ${\bf A}({\bf r})$ which is associated with the magnetic
induction $h({\bf r})$.  The parameters $\alpha$ and $\beta$ have the
usual meaning\cite{Tinkham}. 
Before proceeding any further we must stress a not fully appreciated
fact: Even for small values of $\kappa$ ($\approx 1$),
the magnetic induction is weakly varying in space down to fairly low fields 
($H\approx 0.5H_{c2}$)\cite{Brandt}.
This observation is very important for our purposes 
since, down to $H\approx 0.5H_{c2}$, it is a very good approximation to
consider a uniform magnetic induction [$h({\bf r})=B$] and 
to expand of the order parameter in the following way:
\begin{equation}
\Psi({\bf r})=\sum_{L=0}^{\infty} C_L \frac{1}{\ell\sqrt{2\pi}}
e^{-iL\theta}\Phi_L(r).
\label{expansion}
\end{equation}
In this expansion $C_L \equiv |C_L|e^{i\phi_L}$ are complex coefficients and 
$\frac{1}{\ell\sqrt{2\pi}}e^{-iL\theta}\Phi_L(r)$ are normalized nodeless 
functions that diagonalize, in the symmetric gauge,
the kinetic energy operator
appearing in Eq.\ \ref{G-L} with eigenvalues $\epsilon_L(B)$.
We only consider disk thicknesses smaller than
the coherence length so that the order parameter can be taken constant
in the direction of the field. These eigenfunctions are 
subject to the boundary conditions of zero current through the surface
and the radial part $\Phi_L(r)$ may be found numerically.
(The radial unit is the magnetic length $\ell=\sqrt{e^*\hbar/cB}$).
This expansion captures both the simplicity of the macrovortex (above
$H_{c2}$) when only one $C_L$ is expected to be 
different from zero and the full complexity of 
the order parameter (below $H_{c2}$) when several $L$'s must participate. 

Direct substitution of the expansion \ref{expansion} into Eq.\ \ref{G-L} 
and subsequent numerical minimization of the resulting expression is a daunting
task bound to fail due to the large number of unknown variables 
involved. Instead, it is key to consider expansions 
in restricted sets $\{L_1,L_2,\dots,L_N\}$ of few $N$ components where 
$L_1<L_2<\dots<L_N$. The Gibbs free energy takes the 
following form for each set:
\begin{eqnarray}
G_s-G_n&=&\sum_{i=1}^N \alpha[1-B\epsilon_{L_i}(B)] |C_{L_i}|^2  +
\frac{1}{4}\alpha^2\kappa^2 B R^2 \times \nonumber \\
&&\left[ \sum_{i=1}^N I_{L_i}(B) |C_{L_i}|^4+
\sum_{j>i=1}^N 4 I_{L_iL_j}(B)|C_{L_i}|^2|C_{L_j}|^2 \right.+ \nonumber \\
&&\sum_{k>j>i=1}^N 4 \delta_{L_i+L_k,2L_j}
\cos(\phi_{L_i}+\phi_{L_k}-2\phi_{L_j})  \nonumber \\
&&I_{L_iL_jL_k}(B)|C_{L_i}||C_{L_j}|^2|C_{L_k}| + \nonumber \\
&& \sum_{l>k>j>i=1}^N 8 \delta_{L_i+L_l,L_j+L_k}
\cos(\phi_{L_i}+\phi_{L_l}-\phi_{L_j}-\phi_{L_k}) \nonumber \\
&&\left.I_{L_iL_jL_kL_l}(B)|C_{L_i}||C_{L_j}||C_{L_k}||C_{L_l}| \right]
+(B-H)^2,
\label{LLL}
\end{eqnarray}
where $G_s-G_n$ and $\alpha$ are given 
in units of $H_{c2}^2V/8\pi$ ($V$ is the volume 
of the disk), $\epsilon_L(B)$ is given in units of the lowest Landau level
energy $\hbar\omega_c/2$ (with $\omega_c=e^*B/m^*c$), $R$ is the
radius of the disk in units of $\xi$, and $B$ and $H$ are given
in units of $H_{c2}$. The terms proportional to $\alpha$ contain
the condensation and kinetic energy of the Cooper pairs. All the other terms, 
which are  proportional to $\alpha^2$, account for  the
``interaction'' between Cooper pairs.
There appear four types of these terms: (i) those proportional to
$I_L(B) \equiv \int dr\:r\: \Phi_{L}^4$
reflecting the interaction between Cooper pairs occupying the same
quantum state $L$,
(ii) those proportional to  $I_{L_iL_j}(B)\equiv 
\int dr\:r\: \Phi_{L_i}^2 \Phi_{L_j}^2$
reflecting the interaction between Cooper pairs occupying different
quantum states, and (iii) the ones proportional to  $I_{L_iL_jL_k}(B)\equiv 
\int dr\:r\: \Phi_{L_i}\Phi_{L_j}^2 \Phi_{L_k}$
and proportional to  $I_{L_iL_jL_kL_l}(B)\equiv
\int dr\:r\: \Phi_{L_i}\Phi_{L_j} \Phi_{L_k}\Phi_{L_l}$  which, along
with the phases $\phi_L$, are responsible for the correlation between 
vortices. The non-linear dependence on $B$ of these integrals 
[as well as that of $\epsilon_L(B)$]
comes from the existence of the disk surface.

In order to find the minimum Gibbs free energy for a given set we have to
minimize with respect to the moduli $|C_{L_1}|,\dots,|C_{L_N}|$, the phases 
$\phi_{L_1},\dots,\phi_{L_N}$ of the coefficients, and with respect to $B$. 
The minimum-energy set of components is picked up at the end. The
advantage of doing this selective minimization resides in our
expectation that a small number of components will suffice to describe
the order parameter for the disk sizes considered in the
experiment of Ref.\ \onlinecite{Geim}.
As an illustrative and relevant example 
we consider in the detail the solution with a single component $\{L\}$. 
The energy functional
is invariant with respect to the phase of the only coefficient so 
one can minimize analytically with respect to $|C_L|^2$ to obtain
\begin{equation}
G_s-G_n=-\frac{[1-B\epsilon_L(B)]^2}{\kappa^2 B R^2 I_L(B)}+(B-H)^2.
\label{1c}
\end{equation}
Finally, the minimal value of
$B$ and the minimum Gibbs free energy for each $L$ 
must be found numerically. It
is important to notice that $\alpha$ disappears from the final
expression in Eq.\ (\ref{1c}) which leaves us with
$\kappa$ as the only adjustable parameter when comparing with 
experiments. (This is also true for the more complex cases 
discussed below).   The 2-component $\{L_1,L_2\}$ solutions
can be dealt with in a similar way. The energy functional
is invariant with respect to the phases so one can minimize analytically with 
respect to $|C_{L_1}|^2$ and $|C_{L_2}|^2$ and numerically only with
respect to $B$. The final solutions look generically 
like an $(L_2-L_1)$-vortex ring. For 3-component solutions (two vortex rings)
the energy functional is still invariant with respect to all the three 
phases whenever $L_1+L_3\ne 2L_2$, and, once again, the minimization 
with respect to $|C_{L_1}|^2$, $|C_{L_2}|^2$, and $|C_{L_3}|^2$ can be
done analytically. However, if $L_1+L_3= 2L_2$, 
the two rings have the same number of vortices and their
relative angular positions come into play through the term depending 
on the phases. There is, however, an obvious choice for these phases:
$\phi_{L_1}=0,\phi_{L_2}=0,\phi_{L_3}=\pi$. This choice
gives a negative contribution to the free energy which
reflects the lock-in position between the vortex rings.  
One important fact should be noted now: The components in which the bulk 
Abrikosov lattice needs to be expanded in the symmetric gauge are strongly
overlapping. Depending on the chosen symmetry of the lattice the set
of components is either $\{1,7,13,19,\dots\}$ for the $C_6$ symmetry 
with a vortex at the origin or $\{0,3,6,9,\dots\}$ for the $C_3$ symmetry 
with the center of a vortex triangle at the 
origin\cite{Braverman}.  Consequently, the minimum-energy
solutions for disks are expected to have strongly overlapping components 
which invalidates any perturbative treatment of the terms that contain the 
phases\cite{Palacios}. Moreover, unlike simpler
geometries\cite{Palacios}, there is no direct connection between
number of components and number of vortices. 
This prompts us to seek solutions through
numerical minimization with respect to the moduli and $B$ for the
3-component cases just mentioned, and, for $N>3$, with respect to the moduli, 
the phases, and $B$ whenever the terms involving phases are present. 
Fortunately, for the disk sizes
like the ones used in the experiment of Ref.\ \onlinecite{Geim} 
we will see below that 2 and 3-component solutions suffice to 
capture the relevant physics.

\begin{figure}
\centerline {\epsfxsize=8cm \epsfbox{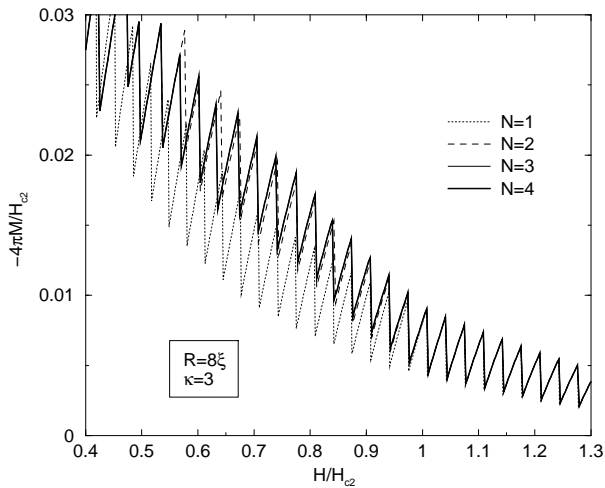}}
\caption{Magnetization as a function of $H$ for a disk of radius 
$R=8 \xi$ and $\kappa=3$. Different line types correspond to different
number of components allowed in the minimization. For this especific set
of parameters 
the plot shows clearly the necessity of considering more than one
component below $H_{c2}$.
However, no appreciable difference can be seen between the traces 
obtained using an expansion with up to $N=3$ and $N=4$ components. }
\label{superdot_acc}
\end{figure}

\begin{figure}
\centerline
{\epsfxsize=5cm\epsfysize=4.7cm\epsfbox{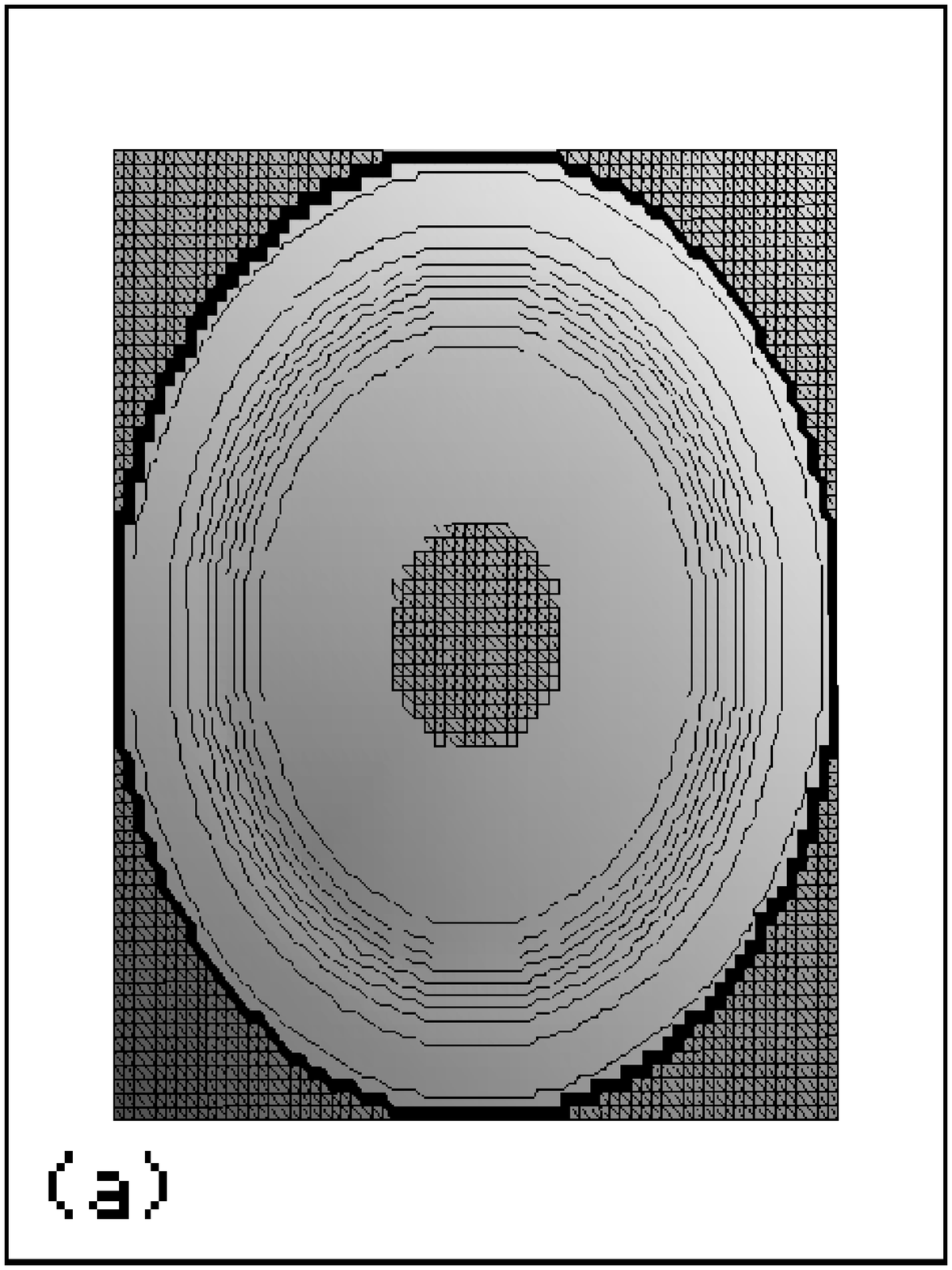}
\epsfxsize=5cm\epsfysize=4.7cm\epsfbox{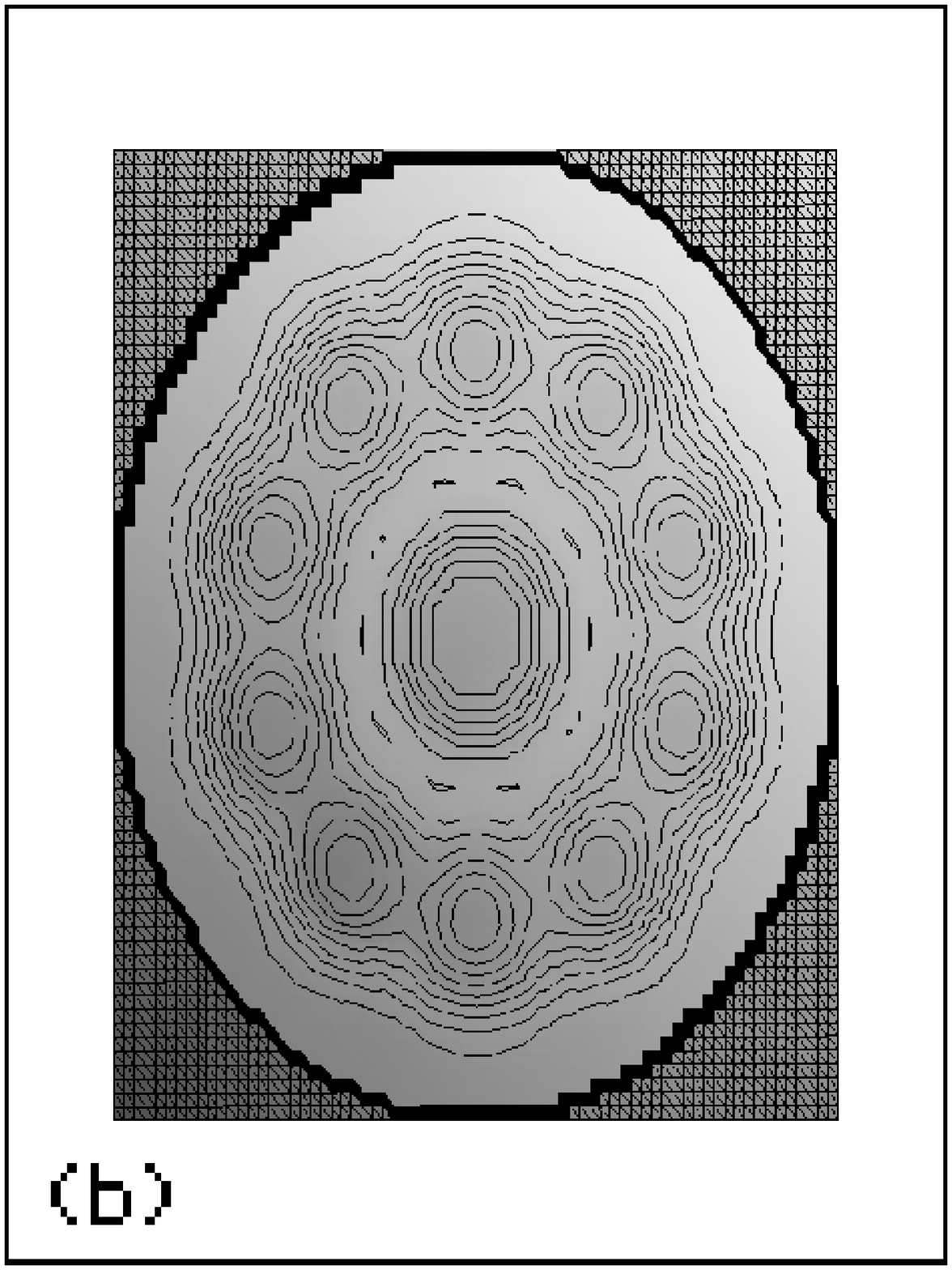}}
\centerline
{\epsfxsize=5cm\epsfysize=4.7cm\epsfbox{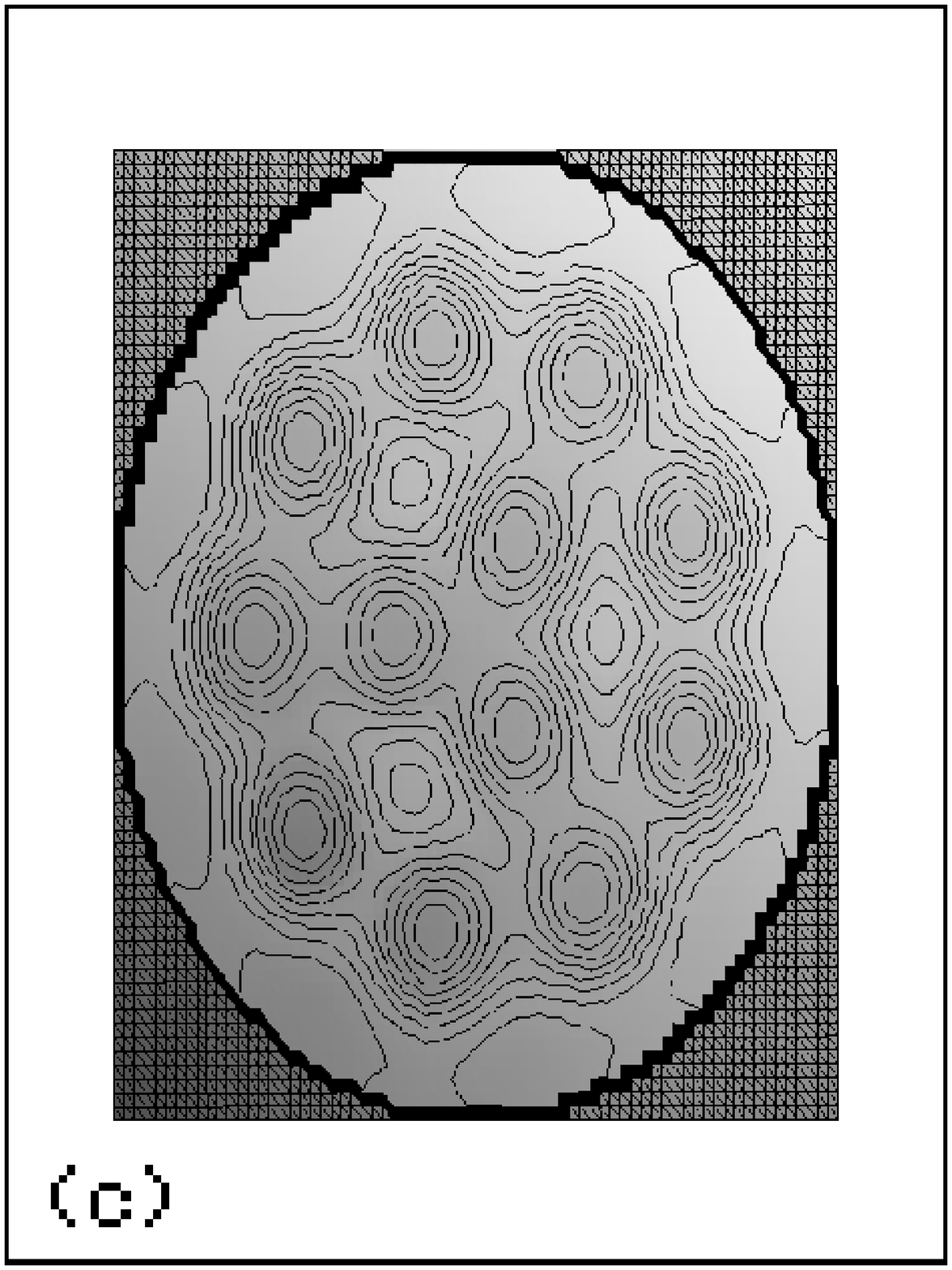}
\epsfxsize=5cm\epsfysize=4.7cm\epsfbox{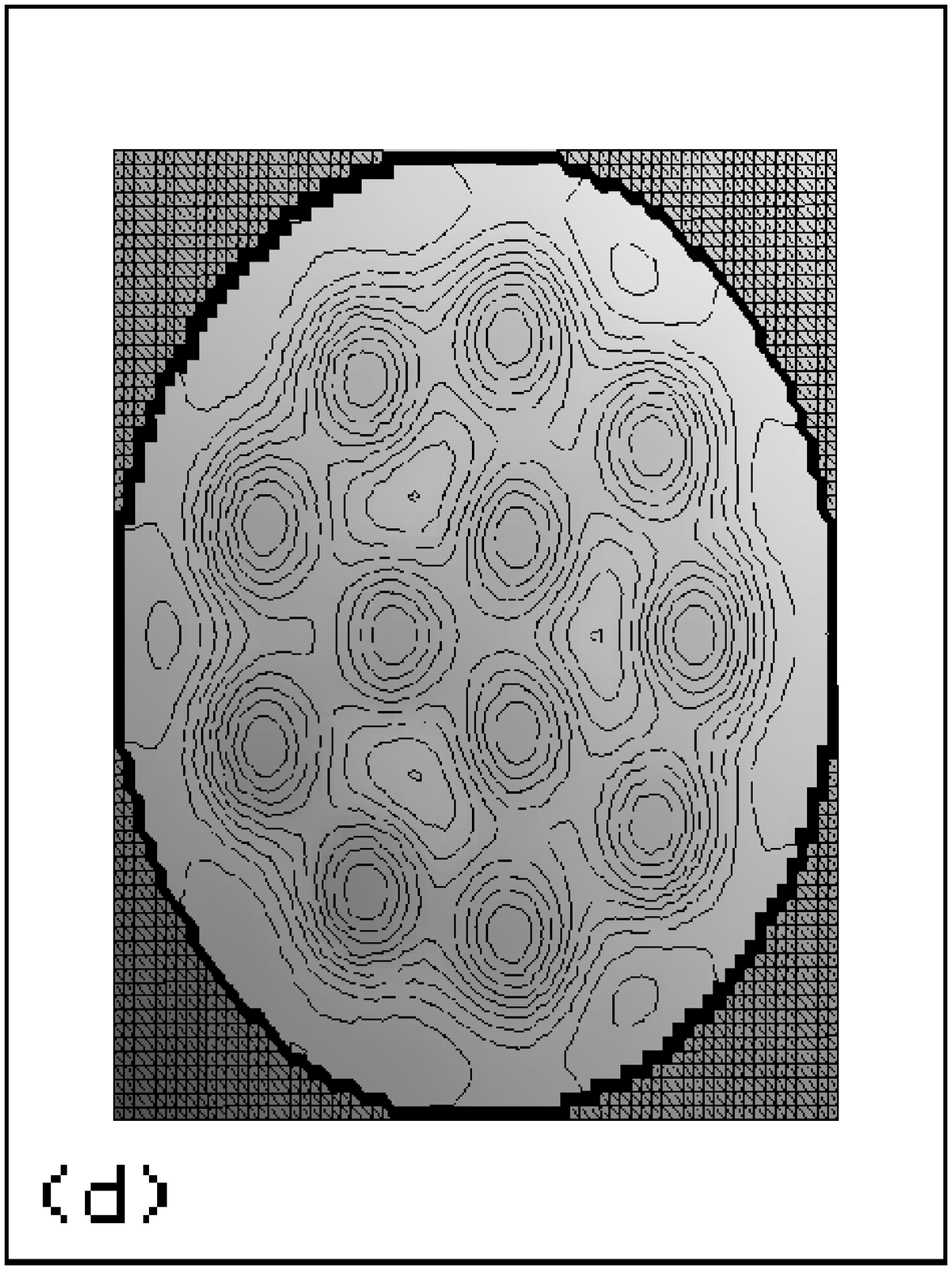}}
\caption{\protect Modulus square of the order parameter 
for an $R=8 \xi$, $\kappa=3$ disk at
$H=0.6H_{c2}$ expanded with up to (a) $N=1$ (minimum-energy set $\{12\}$),
(b) $N=2 \;(\{3,12\})$, (c) $N=3 \;(\{0,3,12\})$,
and (d) $N=4 \;(\{0,3,6,12\})$ components. 
The vortex structure is strongly modified from
$N=1$ to $N=4$ although the total number of vortices 
is always given by the largest $L$ which does not depend on the
the number of components considered. (Only the internal arrangement 
of vortices does.) Allowing  4-component
solutions does not change appreciably the magnetization obtained with
$N=3$ (or even $N=2$) at any value of $H$ (see Fig.\ \ref{superdot_acc}). 
However,  the order parameter is still
modified as can be seen by comparing (c) and (d). 
The addition of a forth component does not increase the number of vortex
rings, but can help fix the relative position of the two existing ones in the
$N=3$ solution.  This modification does not have measurable consequences, 
but illustrates how one of the symmetries  of the triangular 
vortex lattice (the $C_3$) can emerge, bearing confidence in our 
calculations. }
\label{superdot_glass}
\end{figure}

Figure \ref{superdot_acc} shows the magnetization as a function of $H$ for a
disk of $R=8\xi$ and $\kappa=3$. (There is nothing special about these 
parameters, the only purpose of which being to be convenient for the
discussion of the physics we want to bring up).
As indicated in the figure, different types of lines
correspond to the magnetization obtained expanding 
the order parameter with up to $N=1,2,3$, and 4 components. 
A common feature to all curves is that the magnetization exhibits
many first-order transitions. Above $H_{c2}$ the $N=1$ solutions 
suffice to describe entirely the order parameter and magnetization. 
Here the solutions correspond to a giant
vortex which contains $L$ (the quantum number of the single component)
fluxoids. Whenever $L$ changes by one 
the magnetization presents a (non-quantized) jump whose magnitude 
evolves {\em monotonously} with $L$. Below $H_{c2}$ we see that the 
$N=1$ solutions underestimate the correct value of the magnetization.
Allowing more components in the expansion 
has a fundamental effect: It splits the giant
vortex into a complex structure of many single-fluxoid
vortices [for an example see Figs.\ \ref{superdot_glass}(a) to (d)].
This reflects in the magnetization curves by
changing the regular evolution of the magnitude of the jumps into an
irregular one. Whenever a vortex is added or removed from the disk
the symmetry of the new vortex configuration is expected to change
which, in turn, expels the field in a different way. There exist configuration
switches for a given number of vortices, but these changes {\em do not} 
reflect in the magnetization, in contrast to what has been 
suggested\cite{Geim}. When the number of vortices is large enough, 
there are  usually two possible symmetries within
each magnetization step: One with a vortex at the center of the disk
and one without it
which are reminiscent of the $C_3$ and $C_6$ symmetries of the regular
vortex lattice. On top of the many first-order transitions the overall 
slope in the magnetization clearly changes at $H_{c2}$, i.e.,  at
the transition between the giant vortex and the vortex glass structures.
This transition is reminiscent of the second-order transition at
$H_{c2}$ for bulk samples where $M$ vanishes. 

\begin{figure}
\centerline{\epsfxsize=8cm \epsfysize=4cm \epsfbox{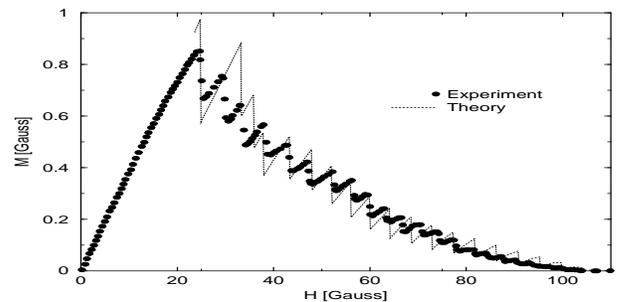}}
\caption{\protect Comparison between experimental data (reproduced from
Refs.\ [10]) and theory using $R=5.25\xi$ and $\kappa=1.2\xi$. Similar
considerations as in Refs.\ [10] have been followed 
for the adjustment of the theoretical curve.}
\label{superdot_exp}
\end{figure}

Finally, we would like to point out that the magnetization measured by
Geim et al. in Ref.\ \onlinecite{Geim} presents features that are in
good agreement with our results despite
of the fact that the disks are made out of a strong type-I material as
Al. In these materials the 
plate geometry can lead to vortex structures being more favorable than
domains\cite{Dolan}. To compare with the experiment, 
we simulate this fact by using a higher value of $\kappa$ than the nominal one. 
In Fig.\ \ref{superdot_exp} we show the data for a disk of nominal
radius $R= 5\xi$, thickness $d=0.6 \xi$,  and $\kappa=0.24$. 
We have obtained a reasonable good
agreement in the number and magnitude of the jumps, and overall shape 
of the magnetization using $R \approx 5\xi$  and $\kappa\approx 1$ 
(the dotted line is a good example). This is consistent
with an effective  penetration length longer than expected
and, possibly, with a coherence length shorter 
than the bulk nominal one.  Although, due to the smallness 
of the disk, it is difficult to point at a
second-order phase transition, the non-monotonous evolution of the
magnitude of the magnetization jumps is notorious over a large 
range of fields which, as
we have shown, is an indication of the formation of vortex glass structures. 
In our approximation the magnetic induction is uniform in space, but,
given the good agreement with the experimental curve,
this seems to be a much less important restriction than considering an
order parameter with a well-defined quantum number $L$\cite{Peeters}.

The author acknowledges enlightening discussions with A. K. Geim and
thanks him for pointing out to the author a very recent related
work\cite{newpeeters}. This
work has been funded by NSF Grant DMR-9503814 and MEC of Spain under
contract No. PB96-0085.

\end{document}